**Title:** the Dead-Alive Physicist experiment: a case-study disproving the hypothesis that consciousness causes the wave-function collapse in the quantum measurement process.

**Authors**: Carlo Roselli and Bruno Raffaele Stella
(*Submitted on 08/06/2020*)

**Abstract**: the aim of this paper is to falsify the hypothesis that the observer's consciousness is necessary in the quantum measurement In order to achieve our target, we propose a variation of the Schrödinger's cat thought experiment called "DAP", short for "Dead-Alive Physicist", in which a human being replaces the cat. This strategy enables us to logically disprove the consistency of the above hypothesis, and to oblige its supporters either to be trapped in solipsism or to rely on an alternative interpretation of quantum mechanics in which the role of the conscious observer has to be reviewed. Our analysis hence provides support to clarify the relationship between the observer and the objects of her/his experimental observation; this and a few other implications are discussed in the fourth section and in the conclusions.

**Comments on the authors**:
Carlo Roselli, independent researcher and also deviser of the experimental setup; e-mail address: <beswick@tiscali.it>; telephone: +39-3335217868; +39-068543272.
Bruno Raffaele Stella, corresponding author, prof. at Dipartimento di Fisica Università Roma Tre and INFN Roma Tre, retired and still active in DESY laboratory in the H1 experiment on quantum cromo-dynamics, e-mail addresses: <bruno.stella@desy.de>, <brunoraffaelestella@gmail.com>; telephone: +39-3332753639; +39-0688642035; ORCID: 0000-0002-5729-8210.

1 **Introduction**

The Copenhagen interpretation of quantum mechanics (QM) does not explain how measurements take place. This gives rise to some thought-provoking demonstrations, usually called "paradoxes", such as the *Einstein–Podolsky–Rosen* (EPR)[1], *Schrödinger's Cat*[2] and *Wigner's friend* experiments, which render questionable the theory's claim to completeness, unless one assumes that consciousness plays a fundamental role in the implementation of the quantum measurement process.

Eugene P. Wigner (1902-1995), following the books published in 1932 and 1955 by the mathematician John von Neumann[3-4] (1903-1957) and a little book published in 1939 by the physicists Fritz London and Edmond Bauer[5], developed an argument in favour of the consciousness assumption, leading to the thesis of the wave-function (WF) collapse at biological-mental level, here more simply called "idealistic interpretation of QM".[1]

This paper proposes an unprecedented version of the Schrödinger's cat and Wigner's friend experiments with the sole purpose of disproving the consistency of the idealistic interpretation of QM. We use the term "idealistic" to refer to the Copenhagen [Niels Bohr (1905-1962)] view of atomic phenomena taken to the extreme. This view asserts that measurements occur, but without providing an explicit definition of the boundary which separates quantum from classical worlds.

This omission gives rise to the so called "measurement problem" with the theory, which weakens its claim to *completeness.*

Indeed, the Copenhagen interpretation is not the only possible interpretation of QM that is subject to this problem which, in more general terms, may be considered as that of defining a satisfactory transition process between micro-systems characterized by quantum state uncertainty and macro-systems obeying the deterministic laws of classical physics. From the 1930's onwards, the measurement problem has been at the centre of a scientific-philosophical debate with the purpose of establishing how (or whether) the collapse of the WF takes place [(see D. Bohm[6-7]; H. Everett[8]; E. Wigner[9-11]; J.A. Wheeler[12]; B.S. DeWitt[13]; W.H. Zurek[14]; G.C. Ghirardi, A. Rimini, T. Weber[15]; G.C. Ghirardi, R. Grassi, P. Pearle, [16]; R. Penrose[17--19]; S. *Hameroff, R. Penrose*[20]; H.P. Stapp[21]; M. Gell-Mann, J. Hartle[22]; C. Rovelli[23]; M. Tegmark[24]; S. Haroche[25]; J.S. Bell[26], and many other illustrious physicists (some of whom are shown in the list of References)]. The debate on this issue has given rise to endless discussions and, so far, there has been a lack of consensus regarding which interpretation might be correct.

Starting from the Copenhagen view, the idealistic interpretation assumes that the observer's consciousness is the only factor which, playing an active role in the measurement process, is capable of determining the transition from the ambiguous realm of potentialities to our familiar and unequivocal realm of actual events. This is the kind of vision we mean when referring to the idealistic interpretation of QM.

The objective of the thought experiment that will be described hereafter is to demonstrate how such an interpretation, also known as "consciousness causes collapse hypothesis" (CCCH), is forced to conclusions incompatible with the assumption that consciousness is necessary for providing a complete explanation of the quantum measurement process.

2 **The DAP experiment**

In this section we are going to propose a rather dramatic experiment, in which an audacious physicist named "**P**" is prepared to give her/his life to science. Imagine **P** inside a sealed room. In a sense, **P** will play the role of Schrödinger's cat as well as the part of Wigner's friend[2] in the original experiments, but – as will be seen – she/he will have a much more important role.

---

[1]There are two main theses arguing that consciousness and quantum mechanical measurement are connected to each other: one thesis (von Neumann, London and Bauer, Wigner, Stapp; see Refs 3-5, 9-11 and 22) holds that the observer's consciousness causes the collapse of the wave function, thus claiming to complete the quantum-to-classical transition, while the other thesis (Penrose; Penrose and Hameroff; see Refs 20 and 21) aims at demonstrating the opposite, i.e. that consciousness emerges from the so called "Orchestrated Objective Reduction".

On the ceiling of the room (figure 1) there is an apparatus, L,[3] programmed to emit at a precise time a photon propagating in a direction along which is located a beam splitter, BS, which forms with it an angle of 45°. Beyond the BS there are two photo-detectors, D and D' with 100% efficiency; D is located along the path of the incident photon, fixed on top of a box and connected to a hammer inside it; under the hammer there is a glass flask containing a lethal dose of gas called "**LGD**"; D' is located along a path which is rotated by 90° with respect to the previous one.

The experiment is planned as follows:
1. L will emit a photon at l2.00 AM hours;
2. **P** has deliberately drugged her/himself one hour beforehand, at 11.00 AM, with a dose of a powerful narcotic, crucial for the experiment, called "**CPB**" (Conscious Perceptions Breaker) and 100% guaranteed to induce a heavy coma-like sleep for two hours;
3. If D is activated, the hammer falls, shatters the flask, the **LGD** is released and subsequently **P** will die.
4. If D' is activated, **P** will be alive conscious at 1:00 PM.

Let us now briefly consider how quantum theory describes the experimental system: the photon WF collides with BS and splits into two component parts, one corresponding to the photon having been transmitted ($T$) and the other to the photon having been reflected ($R$), both with a probability amplitude given by a modulus of a coefficient, which, in our example, is $1/\sqrt{2}$, as shown hereafter:

$$|\psi\rangle = (|\text{photon } T\rangle + |\text{photon } R\rangle)/\sqrt{2} \qquad (1)$$

The combination of these two parts evolves as a linear superposition according to the Schrödinger equation and travels along two different paths, until the instant in which, in agreement with the wave-function collapse (or wave-packet reduction) postulate based on the *Born rule*,[4] one presumes that a measurement process has occurred. This is the instant in which the WF collapses and one only of the possible alternatives becomes real, with probability given by the square of the associated modulus (hence, in our example, probability ½ for both alternatives).

If one assumes, as Wigner, "the existence of an *influence* of the consciousness *on the physical world*" and that "the measurement is not completed until a well-defined result enters our consciousness" (see ref. 11, pp. 181, 187), that is until the WF collapses into either one or the other of its two component parts, then inside the room, as long as **P** is under the **CPB** effect, there is the *linear superposition* described above which, while time is passing, is propagating along the whole macroscopic measurement system up to the scale of **P**'s brain. Subsequently, the superposition will cease to be linear when it reaches **P**'s consciousness.

---

[2]Wigner's friend, here called "**F**", is a physicist left alone inside a laboratory with the task of checking *attentively* whether or not a detector has emitted a flash (has registered the arrival of a photon or not). Wigner is waiting outside and suspects that **F** (as well as all other human beings) may have weird perceptions and be in the superposition of macroscopically distinct states |**F** has perceived a flash⟩ + |**F** has not perceived a flash⟩. Finally, Wigner enters the lab and asks **F** whether or not he perceived a flash. His reply (yes or no) should remove any doubt as to whether the wave-function collapse has occurred. However, Wigner will question whether it is acceptable or not to establish that the collapse into one only of the two possible alternatives is determined by his action (his request and reception of an unambiguous answer). He poses this question since his initial way of interpreting the state of the system gives rise to a rather embarrassing paradox, from which he has three possible ways of escape: 1) accept a relative form of *solipsism*, in the sense that he believes to be, among all living creatures, the only one who has unambiguous perceptions, 2) assume that QM is an incomplete theory, 3) assume that QM is not applicable to human beings; he refutes solipsism and, being a firm supporter of QM completeness, opts for the last solution, assuming that there are beings, at least human beings, endowed with *consciousness* that constitutes an ultimate reality and plays an active role in determining the measurement process by rules that are not susceptible to scientific description.

[3]In such a mechanism a battery is supplying electric power.

[4]This rule, introduced by Max Born in 1926, states that in QM experiments the probability of obtaining any possible outcome, after a measurement/observation, is equal to the square of the corresponding probability amplitude.

Consequently, there are two possibilities or chains of events, which we call "*E(·)*" and which will travel according to the superposition principle until a certain instant of the experiment:
- $E_T$: *T* (transmitted part of the wave function), D registers and triggers the hammer, the flask is shattered, **LGD** spreads in the room, **P** will be *dead* a few seconds after 12:00 PM.
- $E_R$: *R* (reflected part of the wave function), D' registers, the hammer remains hooked, the flask intact, **P** will be *alive conscious* at 1:00 PM.

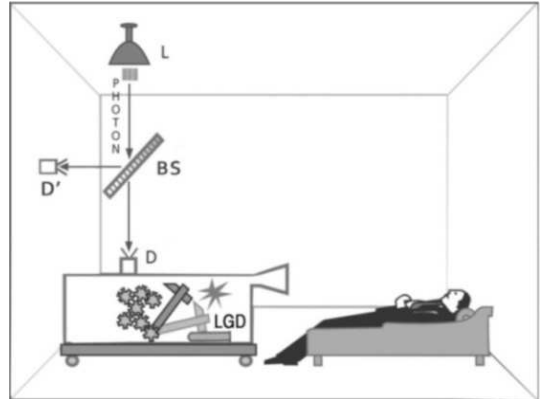

**Fig. 1** Illustration of the experimental room during the time interval between 12:00 AM and 1:00 PM.

It will now be important to analyse the DAP experiment from the perspective of the CCCH. In particular, it is important to try to understand and establish when (at which exact point in time) it makes sense to say that the WF collapse has occurred.

## 3 Formal description of the DAP experiment

In discussing the DAP experiment described in section 2, we will now verify when or whether there are the conditions to bring about the collapse of the WF, $\psi$, of the system.

To this end, during the time interval between 12:00 AM and 1:00 PM, $\psi$ can be formalised as follows:

$$|\psi\rangle = (|T, D \text{ registers}\rangle|\textbf{P} \text{ dead}\rangle + |R, D' \text{ registers}\rangle|\textbf{P} \text{ alive unconscious}\rangle)/\sqrt{2} \qquad (2)$$

The above equation, which entails a superposition of macroscopically distinct states, is not in conflict with the Copenhagen interpretation of QM. The sign "+" between the two incompatible states is a very embarrassing element which does not allow us to assert that **P** is either dead or alive. The CCCH establishes that the superposition can be eliminated only by making on the system an observation aimed at determining whether **P** is dead or alive.

### 3.1 Four remarks regarding the experiment

(i) - The expedient of the **CPB** has a fundamental function in this experiment; in fact, supposing that the **CPB** were not used, according to the idealistic interpretation of QM, **P**'s consciousness would cause the collapse of the WF as soon as **P** has become aware of whether she/he were bound to die or to survive.

(ii) - Equation (2) describes quantum superposition of macroscopically distinct states in sequence, from detectors D and D' to **P**'s brain.

(iii) - Again according to the idealistic interpretation, the collapse of the WF requires that one of its two branches be eliminated by an act of someone's conscious observation.

(iiii) - Human beings have access to their own internal states, perhaps similarly to cats or other animals, but, differently from these, they have the faculty of reasoning, drawing logical conclusions and making *verbal reports*, as it is for example in Wigner's experiment, where his friend is attentive and can provide an answer when questioned about his experience; this last remark, even though not strictly relevant, will make sense in our description.

## 4 The DAP experiment disproves the CCCH consistency

We shall refer to Eq. (2) and imagine an external person named "**W**" (a theoretical scientist supporter of the CCCH) who has in mind to open the sealed room at 1:30 PM. She is convinced that before looking inside it, i.e. during the time interval from 1:00 to 1:30 PM, all the measurement macro-apparatuses. **P** included, are in the entangled superposition of (2), as depicted in the figure below.

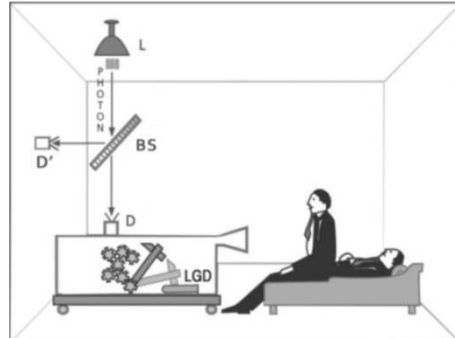

**Fig. 2** From **W**'s point of view, here is depicted the (hypothetical) superposition of macroscopically distinct states of all measuring apparatuses during the time interval between 1:00 and 1:30 PM..

**W**, wearing a gas mask as a precaution, decides to open the room to check whether **P** will be found either *dead* or *alive*:

- **1st** possible outcome : **W** finds **P** *dead* and she asserts that it has been her own act of conscious observation to cause the collapse of the WF. Note that, in this case, **W** performs voluntarily an observation on something that is situated in the physical world outside her mind; here the situation is the same describable in Schrödinger's cat experiment.

- **2nd** possible outcome: **W** finds **P** co*nscious* and asks her/him what happened during the experiment. Promptly **P**, who personally planned the experiment, answers that she/he has certainly been conscious for about the past half hour, starting from 1:00 PM. In this case **W**, being an indomitable supporter of the CCCH, concludes that the appearance of **P**'s consciousness[5] at 1:00 PM collapsed the wave function, eliminating the 'dead' state from the superposition.

However, **P** rejects **W**'s conclusion as logically untenable owing to the considerations that we are going to delineate hereafter.

Since the CCCH implies by definition a causal connection between two phenomena, namely "the observer's consciousness" and "the WF collapse", we will examine whether, in the context of our experiment, there is a way to disprove the former as a causal agent of the latter.

Hence, in order to achieve our goal, we will adopt a line of reasoning putting forward four relevant points, followed by a final argumentation based on mere logic

*First*: we will focus our discussion on the second possible outcome (assuming that there is no more consciousness after death).

*Second*: when we refer to a conscious or self-conscious state, we find it appropriate to distinguish between *intransitive* (passive) and *transitive* (active) self-consciousness.[6]

*Third*: when we consider any QM experiment, particularly Schrödinger's cat, Wigner's friend or similar versions, it is evident that the experimenter, in order to acquire knowledge of the actual outcome, has to perform an intentional observation/measurement that, according to the CCCH, collapses the wave function. In this case, the observer's consciousness is transitive and, more precisely, it plays an active role on the physical world situated outside the mind. Still, it is clear that the above point does not fit the DAP experiment simply because **P**, in view of the introduction of the **CPB** strategy (never thought of before), plays a *passive* role. In fact **P**, as soon as conscious at 1:00 PM, is neither in the same situation of the observer who deliberates to open the Schrödinger's box for verifying the state of the cat nor in the same status of attentiveness of Wigner's friend when checking whether he did perceive a flash or did not.

---

[5] Given the self-referential property of consciousness, the terms "conscious" and "consciousness" will sometimes be replaced in the following by the terms "self-conscious" and "self-consciousness".

[6] See Pereira R. S., "Adverbial Account of Intransitive Self-Consciousness", Abstracta, Vol. 8, Nr 2, PP. 67-77 (2015).

*Fourth*: we point out that 1:00 PM is, by definition, the time in which **P**, as soon as the CPB effect is terminated, undergoes an immediate *intuitive* (non-intellectual) flash that lightens her/his whole being. That is to say that **P** appears intransitively in the state 'conscious'. In particular, at 1:00 PM **P**'s self-consciousness crops up as the "I" that is suddenly disclosed with itself before any objectification resulting from a *subsequent* transitive consciousness involved in cognitive activity. In such a disclosure of consciousness with itself, *subject* (the "I") and *object* ("itself") are indistinguishable and coexist as one and the same accidental entity. By this we mean that **P** appears involuntarily and immediately acquainted with her/himself as an alive conscious being. Thus, this very initial intuitive flash, occurring inside **P**'s body at 1:00 PM, represents an *intransitive* consciousness, that constitutes the surprising (pre-reflexive) realization of her/his immediate existence, to which is expected to follow, at $1:00 + \varepsilon$ PM (with $\varepsilon$ arbitrarily small), a transitive consciousness based on logical thinking, such as when **P**, remembering how the experiment was planned, becomes happily aware of having escaped death, or when she/he deduces that the flask containing the **LGD** must be intact.

*Final and crucial argumentation*: if we again take into account **W**'s claim that the collapse occurred at 1:00 PM, she will be led to conclude that the two events - the appearance of **P**'s consciousness and the WF collapse (thought to be, respectively, the *cause* and the *effect*) - occurred simultaneously:[7] between cause and effect no time has elapsed. But reflecting on this hasty conclusion, we can understand that **W**'s conviction is founded on the wrong impression that at 1:00 PM there are *two distinct events* occurring simultaneously, while at that time, indeed, *only one event* takes place: the fortuitous actualization of the state '**P** alive-conscious'.

This wrong impression, if not grasped as such, gives rise to irrational assertions. In fact, given that, at 1:00 PM, **P** enters the state *'alive-conscious'*, we find it illogical **W**'s belief that **P**'s consciousness causes the collapse of the WF into the same state *'alive-conscious'* that **P** *is entering*.

According to our logic, causality is a relationship through which one event, A (the cause), gives rise to something else, B (the effect). Therefore, the idea (emerging from the above assertion) that the event A gives rise, as effect, to an event which is absolutely not different from A, has to be rejected as an absurdity.

Finally, **P** is now certain that the appearance of her/his consciousness at 1:00 PM could not be the *cause* of the WF collapse. Instead, this is exactly the case in which its appearance must have been the *effect* resulting from the collapse occurred reasonably long before, when D' registered the arrival of the photon.

Summing up, given that the state '**P** alive unconscious' in Eq. (2) changes into the state '**P** conscious' at 1:00 PM, that's it! That is simply the o*utcome* occurring inside **P**'s body when the CPB effect is gone.

**5 Conclusions**

All the above considerations have led us to conclude that, in our experiment, the *collapse* of the WF cannot take place at 1:00 PM. This conclusion implies that the collapse *is independent of the observer's consciousness* and that the CCCH is logically inconsistent, despite a recurring conviction that it is not falsifiable; see for example J. Acacio de Barros and Gary Oas[28].

If our analysis is accepted as well-grounded, a supporter of the idealistic interpretation of QM, like **W**, could still believe that it has been her act of conscious observation to cause the collapse of the WF at 1:30 PM, no matter whether **P** were found dead or alive; but in this case she would be trapped in *solipsism*.[8]

---

[7] The question of causality is problematic, since it requires a distinction between the subjective and the objective aspects of this concept. Causality entails another (arguable) question called "cause and effect simultaneity", which has been discussed and investigated in depth by several philosophers, such as I. Kant, D. Hume, G.W. Leibniz and, recently, by Donald Gillies, Jay F. Rosenberg, Sylvain Bromberger et al.; for a detailed understanding see Buzzoni M.:*The Agency Theory of Causality, Anthropomorphism, and Simultaneity*, section 6, published online: 29 Jan 2015, https://doi.org/10.1080/02698595.2014.979668.

Thus, starting from a rebuttal of solipsism as undesirable in science, one has to rely on an alternative interpretation of QM, in which the role of the conscious observer, who is required in all scientific experiences, is merely relegated to acknowledge the experimental results. We immediately understand that these conclusions have further implications, such as:

(**a**) - the concept of "collapse of the WF independently of consciousness" emerge from the *logical structure* of our thought experiment based on the **CPB** strategy, since it allows to see in a new light the relationship between subject and object of observation (as explained in section 4);

(**b**) - if it were not conceivable an experiment such as the DAP, capable of disproving the CCCH, this latter would still represent a possible and, for a few physicists, even more suitable alternative to the wide range of different interpretations of quantum mechanics.

(**c**) - in the realistic QM theories based on the collapse postulate, *the* boundary between quantum and classical systems should now be reasonably *rescaled down* to the transition point between the quantum system described in (1)] and the initial (uncertain number of) atomic components of the photo-detector with which it interacts at 12:00 + $\Delta t_2$ PM (where $\Delta t_2$ is the travelling time of the photon WF from the source L to D and D');

(**d**) - in the collapse theories *Schrödinger's cat* experiment can no longer be considered a paradox: before opening the box, the cat (as well as **P** in the DAP experiment) is in a statistical mixture of states, 'dead' or 'alive';

(**e**) - the falsification of the CCCH rules out also the hypothesis that the collapse of all the wave-functions involved in our Universe (according to the hypothesis shared by many scientists that consciousness is regarded as an emergent phenomenon) occurred when the first conscious human being appeared in it, thus avoiding to render the big-bang a senseless theory;

(**f**) - One might also argue that the DAP experiment makes the Everett's Many-Worlds Interpretation (or Many-Minds Interpretation) rather problematic [15]. According to this view of QM, no collapse takes place and each of all possible outcomes described by the wave function is physically actualized in some universe. We leave to the readers the task to evaluate the possible implications of this issue, since in one of the two branches of the WF **P** is in the state '*dead*'.

(**g**) - the **CPB** strategy adopted in our experiment can also be applied to cases in which consciousness is present in both branches of the wave function, as described in Wigner's friend experiment. We leave the analysis of this specific issue to further research.

In synthesis, the starting point of our work is that the idealistic interpretation requires the superposition of macroscopically distinct states as well as the conscious perceptive faculty of the observer. This is necessary for consciousness to play a fundamental role in the collapse of the WF. Nevertheless, it is possible to devise at least a thought experiment (e.g. our DAP), which disproves the hypothesis that it is the observer's consciousness that causes the collapse. If this is shared as logically compelling, then one is left with the immediate issue of what the best alternative to the idealistic interpretation should be, and clearly this is an entirely different (and daunting) problem.

However, we feel that the ordinary idea behind our experiment is that there are two ingredients, given by the *wave function* and the *conscious state* of potential observers, which cannot in general be clearly separated, at least in such a way as to make the latter a causal agent in the collapse of the former. If this is true, then a fruitful way to tackle the measurement problem can only be one that treats the above two ingredients in a single coherent framework.

Recent advances in the quantum de-coherence and a re-examination of Everett's Many Worlds Interpretation suggest that such a framework could be constructed entirely within the boundaries of the theory itself; see, for instance, Roland Omnès[29], Maximilian Schlosshauer[30] and

---

[8]There are two different forms of solipsism: one is described in note 2 as a relative solipsism, the other (associated to an inflexible belief in the CCCH) would lead to an absolute solipsism, that is to say "I believe to be the only thinking entity, while an external reality (including my body), being a product of my mind, does not exist.

David Wallace[31], but clearly this is not the only route; see also Bernard d'Espagnat[32] and the very recent works of Art Hobson[33-34].

Furthermore, the DAP experiment could represent a good reason for strengthening some of the actual quantum mechanical spontaneous localization models, where observers have no particular role: we are referring to Ghirardi, Rimini and Weber theory (GRW), to Penrose and to Hameroff-Penrose interpretations, in which the WF is assumed to be as a physical reality and its collapse as an objective dynamical process, that in Penrose's approach is supposed to be induced by gravity.

We would like to close this paper quoting the following words of Steven Weinberg[35, p. 124]:
"*I read a good deal of what had been written by physicists who had worried deeply about the foundations of quantum mechanics, but I felt some uneasiness at not being able to settle on any of their interpretations of quantum mechanics that seemed to me entirely satisfactory*".

**Acknowledgments**

Our special gratitude goes to the late Giancarlo Ghirardi Professor Emeritus of Physics, Università di Trieste, Carlo Rovelli Professor of Physics, Université de Aix-Marseille, Art Hobson Professor Emeritus of Physics, University of Arkansas, Gianni Battimelli Professor of Physics, Università La Sapienza di Roma, Enrico Marchetti, Professor of Economic Policy, Università degli Studi di Napoli Parthenope, and Lorenzo Stella Professor of Physics, Università di Roma 2, for reading and commenting on the manuscript. Finally, we have to thank Susan Beswick for her scrupulous control of the English language of the text.-